\documentclass[12pt]{iopart}

\usepackage{graphicx,iopams,setstack}
\usepackage{cite}
\usepackage{url}
\usepackage{amsfonts}  
\usepackage{mathrsfs} 
\newcommand{\ud}{\,\mathrm{d}} 

\newcommand{\diff}[3][] 
{
  \frac{\mathrm{d}^{#1} #2}{{\mathrm{d} #3}^{#1}} }
\newcommand{\diffp}[3][] 
{
  \frac{\partial^{#1} #2}{{\partial #3}^{#1}} }

\newcommand{\bbN}{\mathbb{N}}

\begin{document}

\title{Quantum Reflection from an Oscillating Surface}

\author{Benedikt Herwerth$^1$, Maarten DeKieviet$^{2}$, Javier Madro\~{n}ero$^3$, Sandro Wimberger$^{1}$}

\address{$1$ Institut f\"{u}r Theoretische Physik, Universit\"{a}t Heidelberg, Philosophenweg 19, D-69120 Heidelberg, Germany}
\address{$2$ Physikalisches Institut, Universit\"{a}t Heidelberg, Im Neuenheimer Feld 226, D-69120 Heidelberg, Germany}
\address{$3$ Departamento di F\'isica, Universidad del Valle, Cali, Colombia}

\date{\today}

\begin{abstract}
We describe an experimentally realistic situation of the quantum reflection of helium atoms from an oscillating surface. The temporal modulation
of the potential induces clear sidebands in the reflection probability as a function of momentum. 
Theses sidebands could be exploited to slow down atoms and molecules in the experiment.
\end{abstract}


\section{Introduction}

Periodically driven quantum system are natural working horses in various fields of physics. For low-dimensional systems, they provide
paradigms of classical and quantum chaos \cite{Izrailev1990299,Breuer1991249,Koch1995289,Buchleitner2002409}.  
From the mathematical point of view, the temporal periodicity allows for the application 
of the Floquet theorem \cite{Breuer1991249} to arrive at effectively time-independent quantum problems which is accessible to analytical treatment, see e.g. 
\cite{Blanes2009151}, or may be simply diagonalized numerically, see e.g. \cite{PhysRevLett.86.3538,PhysRevE.68.056213}. Practically, temporal periodicity is naturally
given by oscillators, e.g. electromagnetic or mechanical waves, providing the external drive.

In this paper we study the reflection of helium atoms from an attractive potential. Since the reflection can then only be caused by quantal effects,
this problem is generally dubbed quantum reflection. Quantum reflections have been experimentally investigated over the last decade by various groups, 
see e.g. \cite{Shimizu2001,DeKieviet2003,Zhao2010,Zhao2011,Dekieviet-ln}, and they are also relevant in the context of hybrid quantum systems \cite{sandro}
involving ultracold atoms bouncing from nanostructures \cite{PhysRevLett.97.093201,Josef2011,javier2011,Josef2012,javier2013}.
As an extension of these experiments with static barriers, we here propose the experimental implementation of a situation 
where the atoms interact with an oscillating barrier \cite{byrd2012}. 
The modulation is assumed to be sinusoidally periodic, which may be realized using a vibrating nanowire \cite{Josef2012} or simply a 
commercially available vibrating membrane in the laboratory \cite{membrane_high_quality}. 

We use a refined numerical method to propagate initial atomic wave packets toward the membrane and investigate their reflection probability. The latter shows
characteristic sidebands induced by the periodic modulation. These sidebands can be clearly resolved for typical experimental parameters. The
controlled addressing of the sidebands allows the experimentalist to engineer the reflection probability and the momentum of the atoms.
Our numerical integration scheme restricts to one spatial dimension, but it is very efficient. This provides a proof-of-principle for the 
possibility to compute precise reflectivities also for higher-dimensional setups. 


\section{Theoretical background and numerical method}
\label{sec:2}

A possible way to describe the long-range atom-surface interaction is given by the phenomenological Casimir-van der Waals potential 
\cite{Shimizu2001,DeKieviet2003,PhysRevA_65_032902}
\begin{equation}
\label{eq:casimirWaals}
 V(x) = -\frac{C_4}{x^3 (x+l)},
\end{equation}
where $x$ is the distance from the atom to the surface and $l$
is the reduced wavelength of the atomic transition. For $x \ll l$, the potential is
proportional to $-1/x^3$, whereas for $x \gg l$ retardation effects are taken into account by the
$-1/x^4$-behavior. The values used throughout this paper correspond to 
the transition wavelength of helium $l = 93 \rm \AA$ and the interaction between helium and a silicon surface $C_{4} = 23.25 \rm eV\AA^4$ \cite{DeKieviet2003,Zhao2010}.

An atom, which is not reflected in the long-range regime and, therefore, approaches the surface up to a distance of some atomic units,
can be subject to inelastic reflections or to sticking processes.
Since we only want to consider the quantum reflection from within
the long-range regime, a regularization of the potential is introduced, which allows particles coming close to the surface
to move further to $x \to -\infty$.
Therefore, the potential is cut at a point $x_0 \gtrsim 0$ and suitably continued for $x \le x_0$.
The reflectivity is then calculated for different values of $x_0$, which allowed to
determine the reflectivity for the quantum reflection by extrapolation to $x_0 \to 0$.
Practically, a parabolic curve was used to continue the potential for
$0 \le x \le x_0$ and a constant potential for $x < 0$. All parameters of this continuation are chosen
such that the potential and its derivative are continuous in $x = x_0$ and in $x = 0$:

\begin{equation} \label{eq:static_potential}
 \fl V_{\rm cont} (x) = \left\{
   \begin{array}{lr}
     V(x), & \mbox{if }  x_0 < x \\
     V(x_0) + V'(x_0) (x-x_0) + \frac{V'(x_0)}{2 x_0} (x-x_0)^2, & \mbox{if } 0 \le x \le x_0 \\
     V(x_0) - \frac{1}{2} V'(x_0) x_0, & \mbox{if } x < 0
     \end{array}
     \right. \,.
\end{equation}
$V$ denotes the Casimir-van der Waals potential of Eq. \eref{eq:casimirWaals} and $V'$ its derivative.
To describe the harmonically oscillating surface, the continued potential $V_{\rm cont}$ is shifted harmonically along the $x$-axis.
We thus use the following time-dependent version of the potential in the Kramers-Henneberger frame of reference \cite{PhysRevLett.21.838}
for our computations:
\begin{equation} \label{eq:non_static_potential}
  W(x,t)  =  V_{\rm cont}(x-d \sin{(\omega t)}) \,.
\end{equation}
Herein $d$ denotes the amplitude of the oscillation and $\omega$ its frequency.

For both the oscillating and the stationary potential $V_{\rm cont}$, the one-dimensional, time-dependent 
Schr\"{o}dinger equation is integrated numerically, using a Gaussian wave packet for the initial state.
The evolution of the initial state is done by a norm preserving Crank-Nicolson scheme \cite{numerical_recipes,pisa}.
Comparing different order approximations of the Hamiltonian with respect to convergence and efficiency, a three-point finite-difference approximation
of the Hamiltonian is finally used in our calculations. Especially when considering a time-dependent system, the three point
approximation turned out to be most effective, because the matrix inversion needed for the time-evolution can be carried out by Gaussian elimination 
\cite{numerical_recipes,pisa}.

When integrating the time-dependent Schr\"{o}dinger equation for a wave packet, artificial reflections arise when the
wave packet hits the boundary of the numerical box. Especially for small values of the connection point $x_0$,
the transmitted part of the wave function is subject to such reflections. Therefore, absorbing boundaries were introduced by
multiplying the wave function $\psi(x)$ by a damping function $f(x)$ after each time step $\Delta t$:
$\psi(x) \to  f(x) \psi(x)$. This damping function is chosen such that it is unity (within numerical accuracy) for $x>0$
and that it falls off smoothly to positive values smaller than unity for $x<0$. Practically, the damping function
\begin{equation}
  \label{eq:damping_function}
  f(x) = \frac{1}{\exp(-\frac{x-a}{\sigma}) + 1},
\end{equation}
is used, where $a$ and $\sigma$ are the two parameters that determine the position and the sharpness of the damping function. 
If $x_b$ is the lower boundary of the numerical box, where the wave function is to be suppressed, $a$ and $\sigma$
were chosen such that the following conditions are fulfilled to minimize artificial reflections, see also Sec.~\ref{sec:3}: 
$f(x_b) = 10^{-8}$ and $|f(x) - 1| < 10^{-16}$ for $x > 0$.

In the case of the {\em time-independent} potential, our approach can be benchmarked with a more straightforward calculation. For this,
the reflectivity is calculated by a numerical integration
of the time-in\-de\-pen\-dent Schr\"{o}dinger equation for a particle with energy $E = \hbar^2 k^2/2m$. The boundary condition for the stationary wave function $\phi$
is chosen such that for $x \to -\infty$ a left-going, transmitted particle is represented by $\phi$. Therefore, a
left-going plane wave is used to define an initial condition $(\phi(x_i), \phi'(x_i))$ at $x_{i} \le 0$, where
the potential (in the continued form) is constant and plane waves are exact solutions to the Schr\"{o}dinger equation.
By numerical integration the wave function and its derivative were obtained at a point $x_f \gg 0$, where the potential
approximately vanishes. The reflectivity is then calculated by matching the wave function at $x_f$ to the sum of an incoming
and an outgoing plane wave:
\begin{eqnarray}\label{eq:time_ind_calc}
 \phi(x_f)  & = & A e^{i k x_f} + B e^{-i k x_f} \\
 \phi'(x_f) &  = & i k A e^{i k x_f} - i  k B e^{-i k x_f}. \nonumber
\end{eqnarray}
We solve Eq.~\eref{eq:time_ind_calc} for the constants $A$ and $B$ and compute the reflectivity as $R=|A/B|^2$.
The adaptive-stepsize Runge-Kutta integrator \emph{odeint} of the \emph{SciPy}-software package \cite{SciPyCite} is used to calculate $(\phi(x_f), \phi'(x_f))$ numerically.

\section{Results}
\label{sec:3}


{\em Static Potential}. Inspired by the experimental setup reported in \cite{DeKieviet2003,DeKieviet2011,Dekieviet-ln}, 
we consider $^{3}He$ atoms with mass $m = 3.01603 {\rm u}$. The reflectivity is first calculated
for the static potential of Eq. \eref{eq:static_potential} by integration of the time-dependent and
the time-independent Schr\"{o}dinger equation. 
 \begin{figure}[t]
   \centering
    \includegraphics[scale = 1.6]{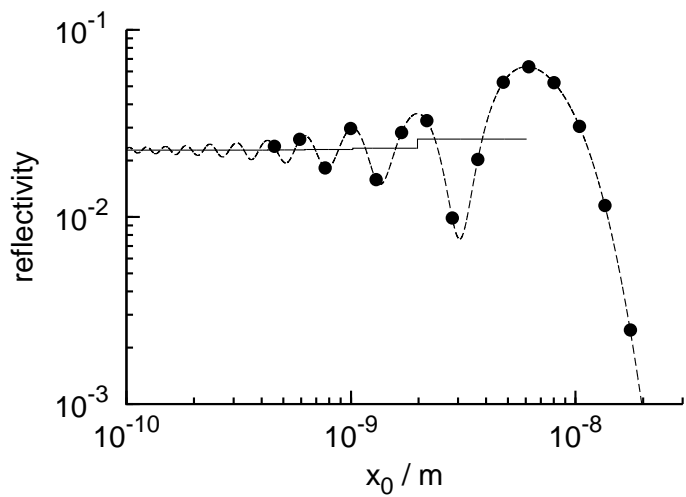}
   \caption{Reflectivity for the static potential as a function of $x_0$. Results obtained by integrating the
     time-dependent Schr\"{o}dinger equation (filled circles) and by solving the stationary Schr\"{o}dinger equation (dashed line)
   are shown. The solid line shows the averages calculated between subsequent maxima by double-geometric averaging. 
 }
   \label{fig:stationary_potential}
 \end{figure}
For the time-dependent integration, the initial state is a Gaussian wave packet
with a mean velocity of $v = -2 {\rm m/s}$ and a relative standard deviation of $\Delta v / v = 3\%$. The initial wave packet is placed
at $x = 4.5 \rm \mu m$. The wave packet is propagated until the reflected part of the momentum distribution becomes stationary. This occurs at times 
$t < 3.4 \rm \mu s$. At $t = 3.4 \rm \mu s$, a Fourier transform of $\psi(x)$ is carried out (efficiently implemented using Fast Fourier Transforms 
\cite{numerical_recipes}) to obtain the
momentum distribution $\psi(k)$. The reflectivity is calculated by integrating the probability distribution
of the reflected particle $|\psi(k)|^2$ in momentum space.

Fig.~\ref{fig:stationary_potential} shows the calculated reflectivity as a function of the connection point $x_0$. We observe an oscillating behavior 
around an asymptotic stationary value, which is reached only in the limit $x_0 \to 0$. The oscillations can be interpreted as a consequence of the ``artificial'' connection
at $x_0$. Since the amplitude of the oscillations decreases as $x_0$ reaches smaller values, the influence of the connection
becomes less and less important for smaller values of $x_0$. The asymptotic reflectivity is best extrapolated 
by taking averages between subsequent maxima as shown in Fig. \ref{fig:stationary_potential} by the solid line.
For $60$ values of $x_0$ between $4 \cdot 10^{-10}\rm m$ and $2 \cdot 10^{-8}\rm m$, the reflectivity is calculated in this way using
the time-dependent and the time-independent method. The values obtained by integration of the time-dependent
Schr\"{o}dinger equation are consistent with those obtained by the time-independent integration within a relative
deviation of the order of one per cent.

\begin{figure}[t]
  \centering
  \includegraphics[scale = 1.8]{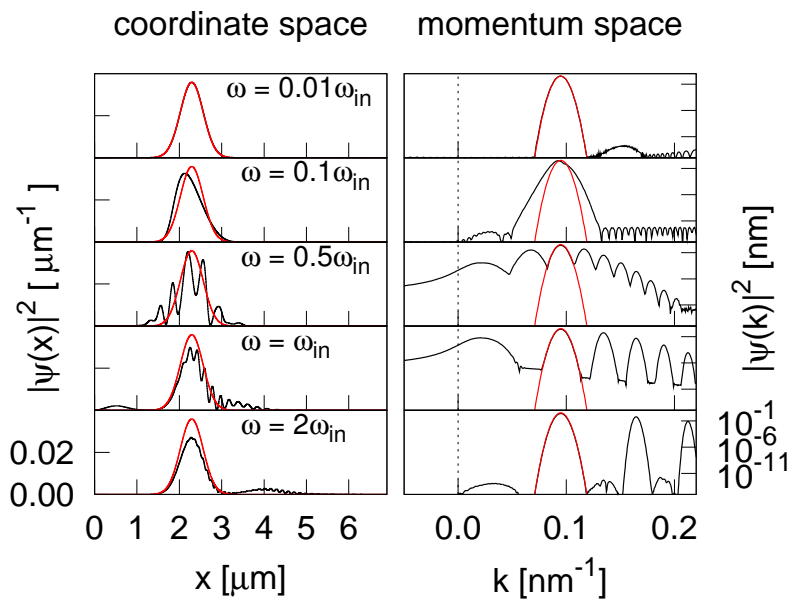}
  \caption{Snapshots of the outgoing wavepackets at $t = 3.4 \rm \mu s$. Coordinate (in linear scale) and momentum distribution (in semilogarithmic scale) 
           for a particle reflected on the oscillating surface (amplitude $d = 4 \rm nm$)
           for different frequencies $\omega$ and fixed $x_0=4\times 10^{-10}$ (black curves). Red curves: corresponding distributions for the static potential.
  }
\label{fig:distr_refl_particle}
\end{figure}

{\em Oscillating Surface.} We now consider the dynamic potential of Eq. \eref{eq:non_static_potential} describing the oscillating surface.
A realistic value of $d = 4 \rm nm$ \cite{membrane_high_quality} is chosen for the amplitude of the oscillation. If $\omega_{\rm in} = E / \hbar$ is the energy of
the incoming particle divided by $\hbar$, the oscillation of the surface is expected to influence the dynamics of
the particle if the frequency $\omega$ is of the order of $\omega_{\rm in}$. For the connection point $x_0 = 4 \times 10^{-10}\rm m$ and a
particle with initial velocity $v=2 \rm m/s$ and $\Delta v / v = 3\%$,
the probability distribution of the reflected particle is calculated in momentum and in coordinate space for different
values of $\omega_{\rm in}$ as shown in Fig. \ref{fig:distr_refl_particle}.

For $\omega = 0.01 \,\omega_{\rm in}$, no effects are visible, whereas for $\omega = 0.1 \, \omega_{\rm in}$,
distortions appear in the distributions.
When $\omega$ reaches $0.5 \, \omega_{\rm in}$, sidebands to the central peak in the momentum distribution become clearly visible.
If the oscillation frequency is increased further beyond $\omega_{\rm in}$, only the right sideband peaks remain. 
This occurs because the left side peak then corresponds to a \emph{negative} momentum, i.e. to a transmitted particle, which is absorbed by the boundary.
On the right hand side in Fig. \ref{fig:distr_refl_particle}, the right peaks moves further away from the central one with increasing frequency. This shows that the energy transfer becomes larger. In the coordinate distributions, significant oscillations imprinted by the moving surface are visible for $\omega \ge 0.5 \, \omega_{\rm in}$.

\begin{figure}[t]
  \centering
  \includegraphics[scale = 1.8]{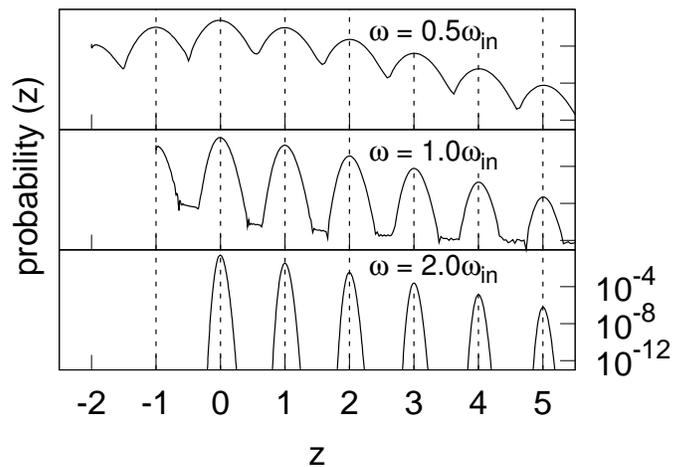}
  \caption{Rescaled and shifted energy distribution $\rho(z)$ of the reflected particle for the same parameters as in
  the previous figure. The distribution is computed
  from the momentum distribution for positive momenta $k$ only (see Fig.~\ref{fig:distr_refl_particle}).
  For each value of $\omega$, the peaks occur at integer values of $z=n \in \bbN$, c.f. Eq.~(\ref{z_variable}).}
\label{fig:distr_energy_refl_particle}
\end{figure}

The energy transfer due to the oscillation of the surface is expected to be quantized by integer multiples of $\hbar \omega$:
\begin{equation}
  E_n  =  \hbar \omega_{\rm in} + n \hbar\omega .
\end{equation}
Because of the nonlinear dispersion relation, 
it is convenient to consider the distribution $\rho'(k)\ud k \equiv |\psi(k)|^2 \ud k$ as a function of the new variable, c.f. also \cite{byrd2012},
\begin{eqnarray} 
\label{z_variable}
  z =  \frac{\frac{\hbar^2 k^2}{2m} - \hbar \omega_{\rm in}}{\hbar \omega} \,,  \\
  \rho'(k)\ud k =  \rho'(k) \frac{m \omega}{\hbar k} \ud z = \rho(z)\ud z \,.
\end{eqnarray}
The peaks in this distribution $\rho(z)$ should now be at integer values of $z$.
As shown in Fig. \ref{fig:distr_energy_refl_particle}, this expectation is fulfilled, which
confirms that the observed side peaks arise from the energy transfer between the particle and the oscillating surface.

\begin{figure}[t]
  \centering
  \includegraphics[scale = 1.3]{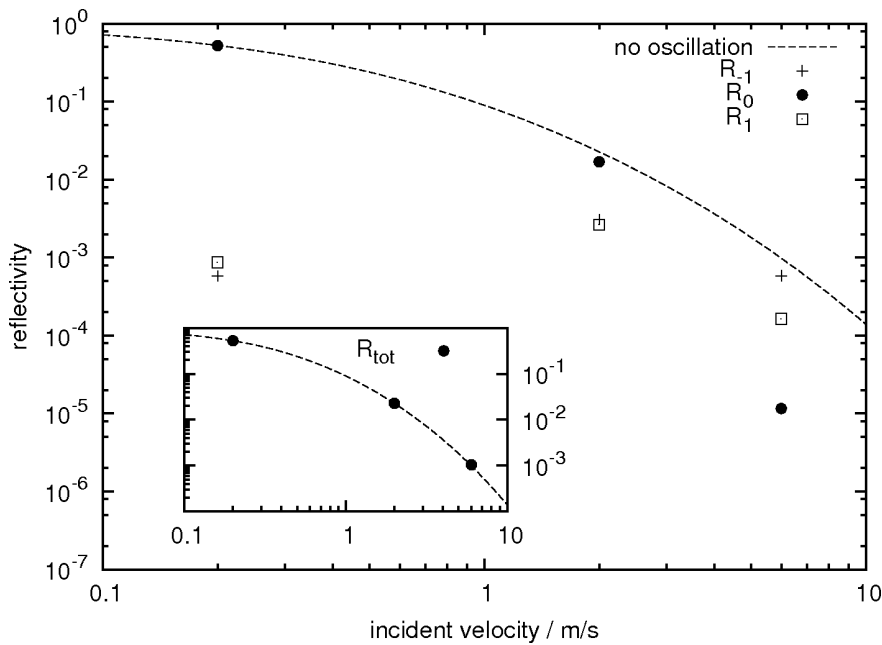}
  \caption{Reflectivity in the presence of the oscillation for $\omega = 0.5 \, \omega_{\rm in}$. The reflectivity for the central peak
  $R_0$ and the two side peaks $R_{-1}$ and $R_{1}$ is shown, together with the reflectivity for the non-oscillating
  surface for comparison (dashed line). To calculate the $R_{-1}$, $R_0$ and $R_1$, $60$ calculations between two subsequent maxima
  in the reflectivity as a function of $x_0$  were carried out in the range $3.7 \cdot 10^{-10}\rm m < x_0 < 7.3 \cdot 10^{-10}\rm m$
  for each velocity. Inset: Reflectivity for the non-oscillating surface (dashed line) and total reflectivity $R_{\rm tot}$ 
  for the oscillating surface (filled circles).}
  \label{fig:refl_with_osc}
\end{figure}

To estimate the reflectivities $R_{n}$ for the different peaks of order $n$ in the momentum distribution in the presence of the oscillation, calculations for
different values of $x_0$ were carried out for $\omega = 0.5 \, \omega_{\rm in}$. As in the case of the non-oscillating surface
(Fig. \ref{fig:stationary_potential}),
the reflectivity for the central peak, $R_{0}$, and for the two first order side peaks, $R_{-1}$ and $R_{1}$, shows an oscillating
behavior as function of $x_0$. Their final value is obtained by taking the average of the reflectivity as a function of $x_0$ between
two subsequent maxima, c.f. the solid line in Fig.~\ref{fig:stationary_potential}. 
The calculated reflectivity for $R_{0}$, which corresponds to an elastic reflection in the laboratory frame, and for $R_{1}$ and $R_{-1}$,
corresponding to an energy transfer of $\pm \hbar \omega$, are shown in Fig. \ref{fig:refl_with_osc} for three
incident velocities.

Our calculations show that the effects caused by the oscillations
of the surface are present over a large range of incident velocities.
For a small incident velocity ($v = 0.2 \rm m/s$), the reflectivity for the central peak dominates. 
For larger incident velocities, the side peaks gain importance. At $v = 6 \rm m/s$, the sidebands even dominate the central peak. 
For $v = 2 \rm m/s$ the reflectivity for the side peaks contributes approximately $13\%$ to the total reflectivity. 
For $v=6 \rm m/s$, it is more likely for the particle to lose than to gain energy ($R_{-1} > R_{1}$). 
This is interesting for practical purposes since it may be used to preferentially slow down atoms or molecules.

As a consistency check, we finally compute the total reflectivity for the three velocities considered in Fig.~\ref{fig:refl_with_osc}. 
This is done by integrating the momentum distribution of the reflected particle around the central and the contributing sidebands.
The calculated reflectivity is compared to the one for the non-oscillating surface in Fig. \ref{fig:refl_with_osc} (dashed line in the main panel and in the inset).
Surprisingly, the obtained total reflectivity $R_{\rm tot}$ corresponds very well to the one of the static case.
For the smaller velocities ($v=0.2 \rm m/s$, $v=2 \rm m/s$), beyond first order sidebands are negligible. For $v=6 \rm m/s$, 
also peaks of higher order become relevant. Here the reflectivity is distributed over many more sidebands. 

Our calculations show that a particle, being reflected on an oscillating surface, may gain energy or transfer energy to the surface,
if the oscillating energy $\hbar \omega$ is of the order of the energy of the particle. In a possible experimental investigation
of this effect, the different scattering channels could be addressed independently and used to improve the precision of the measurement.
Even though oscillating membranes are commercially available \cite{membrane_high_quality}, quantum reflection of matter waves from an oscillating 
surface has not yet been experimentally investigated to the best of our knowledge.

\section{Conclusion}
\label{sec:4}

In this paper we studied the quantum reflectivity of atoms from a periodically oscillating surface for the case of a realistic setup. 
Our numerical method integrates the time-dependent Schr\"odinger equation and controls very well the absorbed part of the incoming wave packet. This allows us
to precisely estimate the quantum reflectivity by an average over the regularization parameter $x_0$ of the potential.
In the computed momentum distribution of the reflected particle sidebands arise, which correspond to an energy transfer of multiples of the 
oscillator energy $\hbar \omega$. The relative contribution of the different sidebands depends strongly on the incoming velocity of the particle.
For large velocities, the first order loss channel $R_{-1}$becomes the dominant contribution. This may be exploited in future experiments to slow down
atomic or molecular beams. For example, the experiment described in \cite{DeKieviet2003} is ideally suited to study quantum reflection from an oscillating 
membrane as proposed here. Although our calculation so far deals with a one-dimensional setup, we are confident that we can generalize our method to a 
two-dimensional geometry. 

\ack
We thank with great pleasure the Heidelberg Center for Quantum Dynamics for funding this project. 
Furthermore, we are grateful for support by the DFG (FOR760), the Helmholtz Alliance Program EMMI (HA-216), and the HGSFP (GSC 129/1).

\section*{References}

\providecommand{\newblock}{}

\end{document}